\documentclass[]{elsarticle}
\usepackage{hyperref,aas_macros,wasysym}
\usepackage[utf8]{inputenc}
\usepackage[greek,english]{babel}
\biboptions{authoryear}
\begin{document}

\begin{frontmatter}
\title{Prior Indigenous Technological Species}
\author{Jason T.\ Wright}
\address{Department of Astronomy \& Astrophysics and \\ Center for Exoplanets and Habitable Worlds\\ 525 Davey 
Laboratory, The Pennsylvania State University, University Park, PA, 16802, USA}
\address{Visiting Associate Professor, Department of Astronomy  \\ Breakthrough Listen Laboratory \\  501 Campbell Hall \#3411, University of California, Berkeley, CA, 94720, USA }
\address{PI, NASA Nexus for Exoplanet System Science}

\begin{keyword}
Astrobiology \sep SETI \sep Venus \sep Mars \sep Solar System \sep Archeology of Space
\end{keyword}

\begin{abstract}

One of the primary open questions of astrobiology is whether there is extant or extinct life elsewhere the Solar System. Implicit in much of this work is that we are looking for microbial or, at best, unintelligent life, even though technological artifacts might be much easier to find. 

SETI work on searches for alien artifacts in the Solar System typically presumes that such artifacts would be of extrasolar origin, even though life is known to have existed in the Solar System, on Earth, for eons. 

But if a prior technological, perhaps spacefaring, species ever arose in the Solar System, it might have produced artifacts or other technosignatures that have survived to present day, meaning Solar System artifact SETI provides a potential path to resolving astrobiology's question.

Here, I discuss the origins and possible locations for technosignatures of such a {\it prior indigenous technological species}, which might have arisen on ancient Earth or another body, such as a pre-greenhouse Venus or a wet Mars. In the case of Venus, the arrival of its global greenhouse and potential resurfacing might have erased all evidence of its existence on the Venusian surface. In the case of Earth, erosion and, ultimately, plate tectonics may have erased most such evidence if the species lived Gyr ago.

Remaining indigenous technosignatures might be expected to be extremely old, limiting the places they might still be found to beneath the surfaces of Mars and the Moon, or in the outer Solar System.

\end{abstract}

\end{frontmatter}

\section{The Search for Other Intelligent Life in the Solar System}

One of the primary open questions in astrobiology is whether life exists or has existed beyond Earth in the Solar System. Mars and the icy moons of Jupiter and Saturn are often cited as perhaps the most likely sites for extant or extinct microbial life to be found, and much effort has been and will be expended to explore them for life. Less commonly mentioned, but also an open question, is whether {\it intelligent} life beyond that on Earth today may exist or have existed in the Solar System. This question poses distinct challenges because technosignatures produced by an intelligent species would be quite different from biosignatures (and might considerably easier to detect.) This is at the heart of the radically different search strategies between Search for Extraterrestrial Intelligence (SETI) and the rest of astrobiology.

SETI typically focuses on interstellar radio signals or other studies of objects beyond the Solar System, however an alternative search avenue has been appreciated for nearly as long: the search for alien artifacts {\it within} the Solar System. This has not only been a topic for science fiction \citep[e.g.\ {\it 2001: A Space Odyssey}][]{Kubrick2001} but in the SETI literature \citep[e.g.][and references therein]{Papagiannis78, Freitas80, Freitas83b, Gertz16}. Indeed, the apparent {\it lack} of such artifacts has been used as evidence that humanity must be the only spacefaring civilization in the Galaxy \citep{hart75}. Despite \citeauthor{hart75}'s claim, we can hardly rule out such artifacts in the Solar System, as demonstrated by \citet{Freitas83a} and \citet{Haqq12}.

In these discussions it is assumed, implicitly or explicitly, that the origin of such artifacts would be not just extraterrestrial \citep[][refer to them as ``Non-Terrestrial Artifacts'' (NTAs)]{Haqq12} but extrasolar.  But if such technology were to be discovered, we should consider the possibility that its origin lies within the Solar System, and potentially on Earth. 

After all, given that the bodies in the Solar System are at least five orders of magnitude closer than the nearest star system, and given that we know that not only are the {\it ingredients} of and conditions for life common in the Solar System, but that one of its planets is {\it known} to host complex life, it is perhaps {\it more} likely that their origin be local, than that an extraterrestrial species crossed interstellar space and deposited it here. At the very least, the relative probabilities of the two options is unclear.

In this paper, I discuss the possibility for such {\it prior indigenous technological species}; by this I mean species that are indigenous to the Solar System, produce technosignatures and/or were spacefaring, and are currently extinct or otherwise absent.

The question of {\it why} this species is not extant in the Solar System is not relevant to much of my discussion, but needs to be addressed at least well enough to establish plausibility for the hypothesis. The most obvious answer is a cataclysm, whether a natural event, such as an extinction-level asteroid impact, or self-inflicted, such as a global climate catastrophe. In the case of a prior spacefaring species that had settled the Solar System, such an event would only permanently extinguish the species if there were many cataclysms across the Solar System closely spaced in time (a swarm of comets, or interplanetary warfare perhaps), or if the settlements were not completely self-sufficient. Alternatively, an unexpected nearby gamma ray burst or supernova might produce a Solar-System-wide cataclysm \citep{Cirkovic16}. Even without a cataclysm, the species may have simply died out, or become permanently non-technological at some point, or \citep[at the risk of committing a ``monocultural fallacy''][]{GHAT1} abandoned the Solar System permanently for some reason.


\section{Prior Art}

The idea that humans are not the first, or only, technological species to arise in the Solar System is very old. In the second century CE Lucian of Samosata wrote (satirically) of intelligent non-human creatures on the Moon in {\selectlanguage{greek}Ἀληθῆ διηγήματα} (True History), and Voltaire (also satirically) wrote of intelligent beings on Saturn in Microm\'egas (1752). Of course, the idea of indigenous Martian civilizations pervades science fiction to the point of clich\'e, but was once also considered at least somewhat seriously in scientific circles, most famously by \citet{lowell1895mars}, but also as recently as \citet{shklovskii1998} (who speculated that the moons of Mars might be artificial).

Since the thorough exploration of the Earth and the robotic exploration of the Solar System has revealed no obvious cities or other signs of non-human civilization on many of these bodies, the idea of other extant technological species has (appropriately) lost much of its scientific currency, but (also appropriately) it has not vanished from the scientific literature \citep[e.g.][who suggest looking for city lights on Kuiper Belt Objects]{Loeb11}.  

Although the possibility of an {\it extinct} indigenous technological species is much harder to foreclose, it is quite hard to find in (and perhaps entirely absent from) the recent peer-reviewed literature. It is even rare in science fiction, though it does appear there occasionally.\footnote{For instance in the ``Engines of Light'' series \citep{macleod2010cosmonaut}; ``Distant Origin'',  (1997, {\it Star Trek: Voyager} Season 3, Episode 23); {\it The Draco Tavern} \citep{niven2007draco}, the Quintaglio Ascension Trilogy \citep{sawyer2007far}, and the Giants series \citep{hogan1977inherit}.} While the applications of this idea in science fiction are usually fanciful, it is unclear to what degree the existence of such species in reality is allowed or disallowed by evidence. 

\section{Terrestrial Origin}

Given that it is known to host complex life, the most obvious origin for a prior species of any sort is Earth.  Archeology and paleontology, having not found evidence for such a prior species or its technology, put strong constraints on when it might have existed and the longevity of its technosignatures. But how long would such evidence last? 

\subsection{Minimum Age: Timescales for Erasure of Technosignatures}

How recently could such a species have existed? \citet{Davies12} and Schmidt \& Frank (2017, in preparation) have thorough analyses of the problem, but for the instant purpose it will suffice to outline it.

The question is not how long the past we might be able to detect the fossil remains of the species---we don't know how to measure intelligence reliably from fossils of bones---but to detect unambiguous technosignatures.

The Earth is quite efficient, on cosmic timescales, at destroying evidence of technology on its surface. Biodegredation can destroy organic material in a matter of weeks, and weathering and other forms of erosion will destroy most exposed rock and metals on a timescale of centuries to millennia, if human activity does not erase it faster. At the very longest, some especially large and durable structures in the right environments---the Great Pyramids, for example---might last for tens of thousands of years.

Some signatures will last longer. Many forms of preservation---tar, ice, isolated caves in arid regions---might work over tens of millenia, but will still fail on longer timescales. Fossilization may work on physical pieces of technology on longer timescales. Mining leaves long-term scars on the terrain, and also depletes an area of ore, and so a global civilization that valued, say, coal or iron might leave depletions of those resources that would be obvious to geologists for much longer, perhaps millions of years. Fossilization or preservation in amber can preserve some records of living creatures for hundreds of millions of years, but this represents a small fraction of the Earth's surface, and will not obviously preserve any technosignatures.

Nuclear activities will create not just unnatural, short-lived isotopes, but  unnatural isotope ratios from the stable daughter products of decay that might be obvious essentially forever.

Humanity seems to have had a sufficient impact on the Earth that it has created an unambiguous geologic record of its technological activities \citep[the ``anthropocene'', e.g.][]{zalasiewicz2011anthropocene}. A prior species with a similar effect would thus probably have been noticed in the geological record.

On a timescale of hundreds of Myr or Gyr, however, plate tectonics will subduct almost all evidence for technology with the crust it sits upon, erasing it from the surface entirely. The parts of the surface that escape subduction also change substantially on tectonic timescales, so regions that are easily accessed today might have been practically inaccessible at the time a prior species existed (under miles of ice, for instance), and so show few or no signs of their technology. 

The present-day detectability of technosignatures is thus a strong function of their age. Historical records would reveal any such species less than a few thousand years old. Archeology would reveal technosignatures less than a few tens of thousands of years old.  The geological record of the past few hundred million years might show a distinct layer if the technology had a widespread geological effect, as ours does.  But beyond this, on Gyr timescales the isotopic or chemical signatures of technology on the Earth's surface might be quite subtle, and possibly misinterpreted as natural, or there may be nothing to be found at all. 

\subsection{Maximum Age: Compatibility with the Fossil Record}

The next question is then how {\it long ago} could such a species could have existed. Complex life has been common on Earth since the Cambrian ``explosion'' around 540 Myr ago; before this the fossil record contains only much simpler organisms, such as single-celled species and their colonies. We would then expect that any prior intelligent species to be no older than this event.

But we should perhaps keep an open mind about even this conclusion. We associate intelligence with complex life that develops a nervous system using biological mechanisms that evolved in the Cambrian explosion, but perhaps colonies of single-celled organisms were able to organize in complex ways prior to this that achieved the same effect. Alternatively, perhaps there was a prior ``explosion'' of biological complexity in Earth's more distant past, farther back than the fossil record is reliable, or that produced a form of complex life that leaves little or no fossil record. A planet-wide cataclysm (perhaps the same one that extinguished our hypothetical species) might have destroyed all such prior complex life, forcing the biosphere to ``start over'' with the few single-celled species that survived \citep[perhaps on a rock ``lifeboat'' ejected during the offending asteroid impact,][]{Lifeboat}.  The first generation of complex life would then be difficult to find, evidence for it existing only in the most ancient rocks, if anywhere.  
 
\section{Other Origins}

Present-day Venus would seem to be a terrible candidate for a technological species, with a surface temperature over 700K, although when it comes to alien life we should keep an open mind about even this.  At any rate, radar mapping of its surface means that we can be all but certain that it has no technological species on its surface {\it today} that generates large, obvious topographical anomalies.

But the Venusian surface, thick atmosphere, and intense greenhouse may not be ancient \citep{VenusSurfaceAge}. There may have been episodes of catastrophic resurfacing \citep{VenusResurface}, destroying all evidence of a prior biosphere. Ancient Venus would have had a much thinner atmosphere when the Sun was significantly fainter, and the surface may have been habitable \citep{Way16}.

Ancient Mars likely had liquid surface water and may have been habitable. \citep{Masursky77,Pollack87,Craddock02,Squyres04,Fassett08,Ramierez14,WaterGaleMars} As such, it is often considered the most likely place to find evidence for extraterrestrial life in the Solar System. After Earth, it is thus also perhaps the most likely host for a prior indigenous technological species. Next, the icy moons, and even the asteroids, make obvious sites where an intelligent species may have arisen. In most of these cases, it is possible that evidence for that or related life may still exist {\it in situ} (indeed, discovery of such life is a major priority for astrobiologists).

In all cases, it should be noted, the life need not have had an independent genesis to Earth's life. Lithopanspermia \citep[the spread of life among bodies in space via rocks ejected from impacts,][]{Lithopanspermia, Worth13} may have seeded the entire Solar system with life from a common abiogenesis (which may or may not have been on Earth).

\section{Technosignatures Throughout the Solar System}

While all geological records of prior indigenous technological species might be long destroyed, if the species were spacefaring there may be technological artifacts to be found throughout the Solar system. 

\citet{Haqq12} discuss some of the difficulties in assessing the completeness of our search for Solar System artifacts, concluding that our completeness is very low. Given that artifacts from a prior indigenous technological species might be very old, it is possible that any remaining technosignatures would be very difficult to find.

Many hypotheses for why alien artifacts might exist in the solar system involve a system that monitors the Solar System or announces itself \citep[a ``beacon,''][and references therein]{Burke-Ward00,Haqq12}. Indeed, the Breakthrough Listen radio SETI program includes asteroids on its target list \citep{Isaacson17} following the suggestion by \citet{Gertz16}.

In the case of a prior indigenous technological species, the artifacts might have had totally different purposes, such as asteroid mining operations or settlements on other planets and moons. Such structures would be expected to fall into disrepair, especially if its creators are absent. Given the large amounts of time since a prior indigenous technological species could have arisen on Mars, Earth, or Venus, it is possible that any artifacts from such a species have long ago become inoperative. This restricts the opportunities for their detection.

Consider technosignatures on Mars. Unlike Earth, its surface is ancient \citep{Farley14} and so might host very old artifacts. It does experience significant erosion due to dust and wind, however, so artifacts might need to be below the surface to have survived. In particular, large structures or artifacts might have become buried under dust and eventually protected from erosion (though not large impacts). As such, it is unlikely that artifacts would be obvious from space imagery, or even from the sort of shallow probing performed by the various Martian rovers. 

Artifacts on the rocky moons \citep[including ours][]{Davies13} or on asteroids need not worry about erosion from wind or rain, but micrometeorites (and larger bodies) will eventually ``garden'', erode, and destroy them on timescales longer than Myrs, depending on their size and durability \citep{Szalay16}.  Structures buried beneath surfaces, however, might survive and be discoverable as long as they do not suffer a collision so severe that their artificial nature is obliterated (merely destroying them would render them nonfunctional, but they might still be recognizably technological).  We might conjecture that settlements or bases on these objects would have been built beneath the surface for a variety of reasons, and so still be discoverable today.

The surfaces of icy moons are significantly younger \citep[e.g.][]{Zahnle98} and so will likely not be good sites for preserving very old technosignatures. 

Free-floating objects not designed to last for Myr--Gyr in the Solar System will suffer from several problems. If the artifacts are inert, lacking propulsion, they will be be subject to the Solar System's dynamical chaos and solar radiation pressure. Most of the few stable orbits that do exist in the inner Solar System are not empty, and so even objects in long-term stable orbits will suffer from collisions. The present lack of asteroids in much of the inner Solar System illustrates how these processes can remove objects on Gyr timescales. The lifespan of free-floating artifacts will thus be sensitive to their size and mass (which determines their sensitivity to radiation pressure), their level of collisional shielding, and their location, all of which affects their collision rate. Artifacts in the Kuiper Belt  might survive long enough for eventual discovery.

\section{Conclusions and Future Work}

Prior indigenous technological species may have existed in the Solar System. Given the signatures humanity's technology has already imprinted on our future geological record, we might expect such a prior species on Earth to have made a similar impact. The study of the oldest rocks on Earth for technosignatures---including unnatural isotope ratios, synthetic elements, or evidence of mining---might thus be a fruitful exercise. It also may be that any such species that arose on Earth or Venus have left no trace that we can ever discover {\it in situ}.

If such a species were spacefaring or arose elsewhere, however, more opportunities for its discovery exist. It might have left more unambiguous technosignatures in the form of artifacts beneath the surface of Mars, the rocky moons and asteroids, or in orbit in the outer Solar System where they could be discoverable. 

Such discoveries might occur using the tools of the burgeoning field of the archeology of space \citep[e.g.][]{gorman_archaeology_2005}, which includes searching for, finding, and interpreting {\it human} artifacts in space.\footnote{The field is broader than this, and also includes understanding and preserving sites on Earth associated with space travel.}  Such work includes the re-discovery and identification of lost probes and other space-borne human artifacts either for forensic purposes \citep{Abdrakhimov11,Tao16,Wagner17}, or even accidentally \citep{Denisenko13}. 

Perhaps more likely, imagery and subsurface radar used to study the geology of planetary surfaces might reveal traces of buried structures or other artifacts.  Photometry and spectra of asteroids, comets, and Kuiper Belt Objects might reveal albedo, shape, rotational, compositional, or other anomalies because the targets host, or are, artifacts.

Further work to foreclose or constrain the possibility of prior indigenous technological species would come from a firm understanding of the conditions in Venus's pre-greenhouse past, and an analysis of the of pre-Cambrian-explosion Earth that showed it must have been the first such event on Earth.

Further theoretical work to foreclose or constrain the presence of free-floating artifacts in the Solar System includes determining where such objects might safely orbit for Gyr, especially assuming they are unpowered, and what their collisional timescales would be for a reasonable range of parameters of shielding, size, and mass.  A similar analysis of the effects of collisions would be valuable for structures on or just beneath the surfaces of Mars, the Moon, or asteroids. This might constrain the maximum age a structure of a given size or material could be without having been obliterated by impacts.

\section*{Acknowledgements}

I thank Michael Oman-Reagan, Andrew Shannon, Jenn Macalady, Steinn Sigur\dh sson, Marshall Perrin, James Guillochon, Eric Huff, Shawn Domagal-Goldman, David Grinspoon, Seth Pritchard, Geoff Marcy, Lindy Elkins-Tanton, Ramses Ramierez, Natasha Batalha, Emily Mitchell, Andreas M\"uller, and Julia Kregenow for suggestions.

The Center for Exoplanets and Habitable Worlds is supported by the
 Pennsylvania State University, the Eberly College of Science, and the
 Pennsylvania Space Grant Consortium. This research was partially supported by Breakthrough Listen, part of the Breakthrough Initiatives sponsored by the Breakthrough Prize Foundation,\footnote{\url{http://www.breakthroughinitiatives.org}}.

\section*{References}


\end{document}